\documentclass[apjl]{emulateapj}

\def\Lsun{\hbox{\it L$_\odot$}}

\def\Msun{\hbox{\it M$_\odot$}}
\def\Minit{\hbox{\it M$_{\rm initial}$}}

\def\Myr{\hbox{\it Myr}}

\def\kms{\hbox{km$\,$s$^{-1}$}}

\def\simgr{\mathrel{\hbox{\rlap{\hbox{\lower4pt\hbox{$\sim$}}}\hbox{$>$}}}}

\def\lbv{\hbox{LBV~1806$-$20}}

\shorttitle{LBV 1806$-$20}
\shortauthors{FIGER, NAJARRO, \& KUDRITZKI}

\begin{document}

\title{THE DOUBLE-LINED SPECTRUM OF LBV 1806$-$20
\footnote{Data presented herein were obtained at the W.M. Keck
Observatory,  which is operated as a scientific partnership among the
California Institute of Technology,  the University of California and
the National Aeronautics and Space  Administration.  The Observatory
was made possible by the generous financial support of the W.M. Keck
Foundation.}}

\author{
Donald F. Figer\altaffilmark{1}, Francisco Najarro\altaffilmark{2}, Rolf P. Kudritzki\altaffilmark{3}}

\email{figer@stsci.edu}

\altaffiltext{1}{Space Telescope Science Institute, 3700 San Martin Drive, Baltimore, MD 21218; figer@stsci.edu }
\altaffiltext{2}{Instituto de Estructura de la Materia, CSIC, Serrano 121, 29006 Madrid, Spain }
\altaffiltext{3}{Institute for Astronomy, University of Hawaii, 2680 Woodlawn Drive, Honolulu, HI 96822}

\begin{abstract}
Despite much theoretical and observational progress, there is no known 
firm upper limit to the masses of stars. Our understanding of the interplay between the immense
radiation pressure produced by massive stars in formation and the opacity of infalling
material is subject to theoretical uncertainties, and many observational claims
of ``the most massive star'' have failed the singularity test. \lbv\ is a particularly
luminous object, L$\sim$10$^6$~\Lsun, for which some have claimed very high mass estimates 
(\Minit$>$200~\Msun), based, in part, on its similarity to the Pistol
Star. We present high-resolution near-infrared spectroscopy of \lbv, showing that
it is possibly a binary system with components separated
in velocity by $\sim$70~\kms. If correct, then this system is not the most
massive star known, yet it is a massive binary system. We argue
that a binary, or merged, system is more consistent with the ages of nearby stars in the
\lbv\ cluster. In addition, we find that the velocity of V$_{\rm LSR}$=36~\kms\ is consistent
with a distance of 11.8~kpc, a luminosity of 10$^{6.3}$~\Lsun, and a system mass of $\sim$130~\Msun.
\end{abstract}


\section{Introduction}
Massive stars are key ingredients and probes of astrophysical phenomena on all 
size and distance scales, from individual star formation sites, such as Orion, 
to the early Universe during the age of reionization when the first stars were born. 
As ingredients, they control the dynamical 
and chemical evolution of their local environs and individual galaxies through their influence on the 
energetics and composition of the interstellar medium. They likely play an
important role in the early evolution of the first galaxies, and there is
evidence that they are the progenitors of the most energetic explosions in the Universe, seen
as gamma ray bursts. As probes, they define the upper limits of the star formation process 
and their presence likely ends further formation of nearby lower mass stars. 
They are also prominent output products of galactic mergers, starburst galaxies, and AGN. 

Despite their central role in these varied topics, 
there is no known firm upper limit to the maximum stellar mass. Such a basic 
quantity escapes both theory, because of the complex interplay between radiation 
pressure and opacity, and observation, because of incompleteness in surveying the 
Galaxy along the plane and the dearth of resolved starburst clusters with 
masses $>$10$^4$~\Msun. The most promising cases are made recently by 
\citet{fig03} and \citet{wei04}, who give observational
evidence for an upper mass cutoff of $\sim$150~\Msun\ in the Arches cluster and R136,
respectively. Finding even a single star with a mass significantly beyond this
limit would single-handedly change our understanding of the maximum mass a star
may have. 

Of course, each ``discovery of the most massive star'' must face the crucial test of
singularity. Some such claims have not passed this test, with subsequent observations
revealing that the object is actually composed of multiple stars.
For instance, \citet{wei91} 
find that R136 is actually a massive cluster of stars, rather than a 
supermassive single star with M$\sim$250-2,000~\Msun\ \citep{fei80,pan83}.
Likewise, \citet{dam00} claim that $\eta$~Car is binary, with present-day masses of $\sim$70~\Msun\ and
$\sim$30~\Msun, rather than a single star with M$\simgr$100~\Msun. 
 
Some of the most massive single stars known are near the Galactic Center. 
\citet{fig98} identify the Pistol Star, in the Quintuplet Cluster, as one of the most massive 
known, with an initial mass of $\sim$200~\Msun, 
and show that the star is single based upon their Keck speckle data 
and spectra; the former reveal that the star is single down to a 
projected distance of 110~AU~(14~mas), while the latter do not show an 
obviously composite spectrum. \citet{fig99a} identify a near-twin 
to the Pistol Star, FMM362, less than 2$\arcmin$ from the Pistol Star, and \citet{geb00} 
demonstrate that it had the same spectroscopic and 
photometric characteristics as the Pistol Star, at the time of the observations. Presumably,
FMM362 is nearly as massive as the Pistol Star. 

\citet{kul95} identify a luminous source near the soft gamma ray repeater, SGR~1806$-$20.
\citet{van95} determine that this source has characteristics similar to Luminous Blue Variables (LBVs), i.e.\ it
has L$\simgr$10$^6$~\Lsun\ and a K-band spectrum indicating a spectral type in the
range O9-B2, leading them to propose that it is an LBV-candidate. 
\citet{eik01,eik04} estimate the properties of the star, finding 
a luminosity, and thus mass, that is at least comparable to that of the Pistol Star.
\citet{eik04} obtained high resolution images at Palomar, showing that the
star is single down to about 900 AU. Note that the putative binary
components of $\eta$~Car have a maximum separation of ~30~AU \citep{dam00}. So, it is possible that \lbv\ 
is multiple, but it is certainly not a ``cluster,'' in the traditional sense.
Even the most compact cluster in the Galaxy, the Arches cluster, has a
half-light radius of $\sim$40,000~AU \citep{fig99b}. 

In this Letter, we present high-resolution near-infrared spectra of \lbv\ showing
that the helium absorption lines are double, suggesting that the star may be binary. 

\section{Observations and Data Reduction}
The data were obtained at Keck on 22 June 2003, using NIRSPEC \citep{mcl98}. \lbv\ was observed
at high resolution (R$\sim$26,500=$\lambda/\Delta\lambda_{\rm FWHM}$) from 1.68~\micron\ to 2.25~\micron\
in three grating settings, and at
low resolution (R$\sim$2,500) from 2.7~\micron\ to 4.2~\micron\ in two grating settings. 
We obtained two exposures per grating setting, with the telescope nodded along
the direction of the slit inbetween exposures. 
The exposure times were 200 seconds, 100 seconds, and 30 seconds, for the high resolution
short-wavelength, long-wavelength, and low resolution pointings, respectively.

Quintuplet Star \#3  (hereafter ``Q3''), which is featureless in this
spectral region  \citep[Figure~1]{fig98},
was observed as a telluric standard  \citep{mon94}. Arc lamps containing Ar, Ne, Kr, and
Xe, were observed to set the wavelength scale. In addition, a continuum
lamp was observed through an etalon filter in order to produce an
accurate wavelength scale between arc lamp lines and sky lines
(predominantly from OH). A quartz-tungsten-halogen lamp was observed to provide a ``flat''
image which was divided into the background-subtracted target images. 

\begin{figure}
\plotone{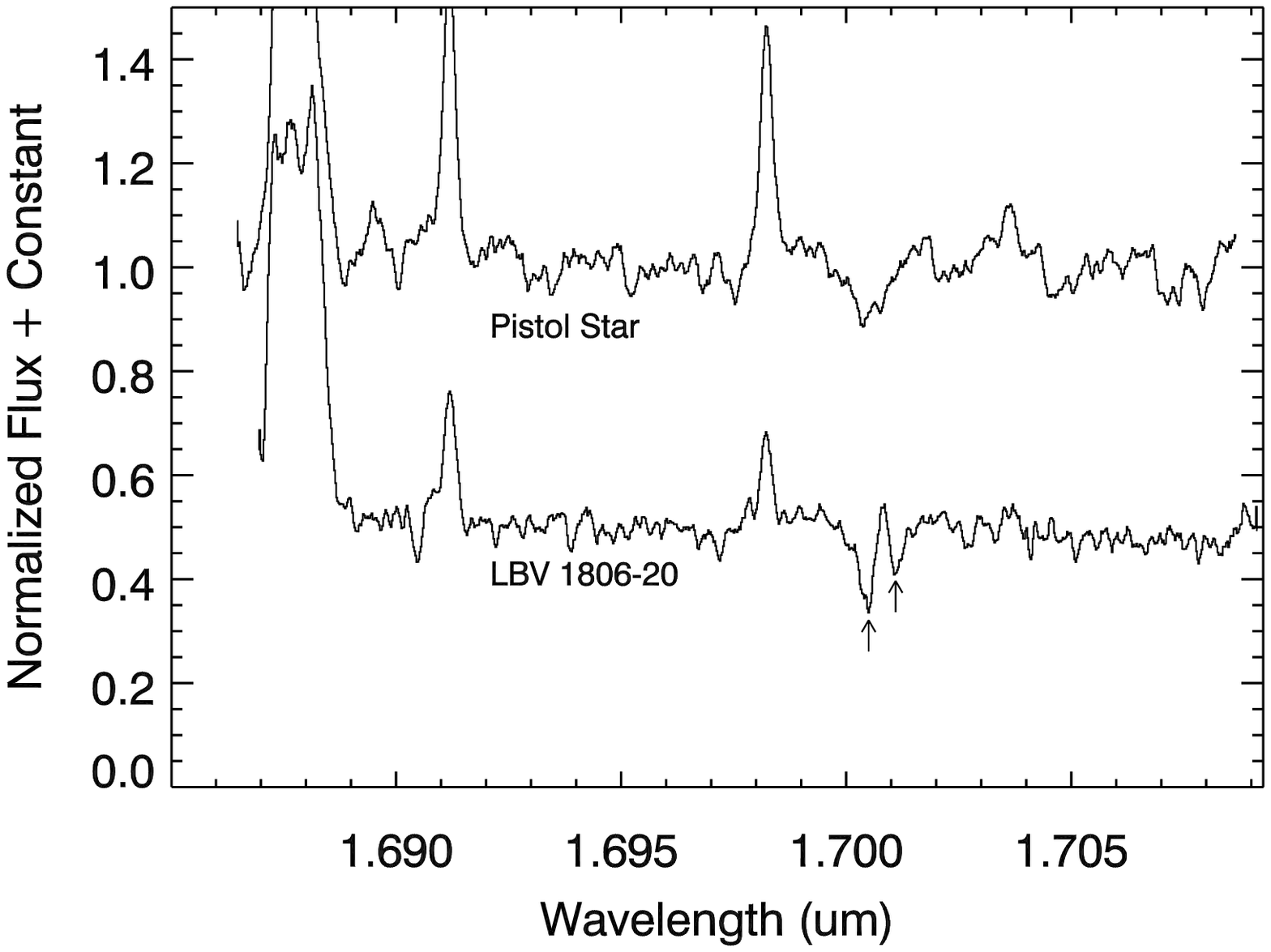}
\caption
{\label{fig:1700}Spectra of the Pistol Star and \lbv\ near the 1.70~\micron\ \ion{He}{1} transition. 
The feature in the spectrum for \lbv\ is double (marked by arrows), with the blueward
component having a P Cygni signature. The broad feature on the left at 1.688~\micron\ is attributable to
an \ion{Fe}{2} transition. The two features at 1.691~\micron\ and 1.698~\micron\ are due to \ion{Si}{2}.
The weak absorption features near 1.6905~\micron\ and 1.6960~\micron\ are due to telluric absorption.
}
\end{figure}

\begin{figure}
\plotone{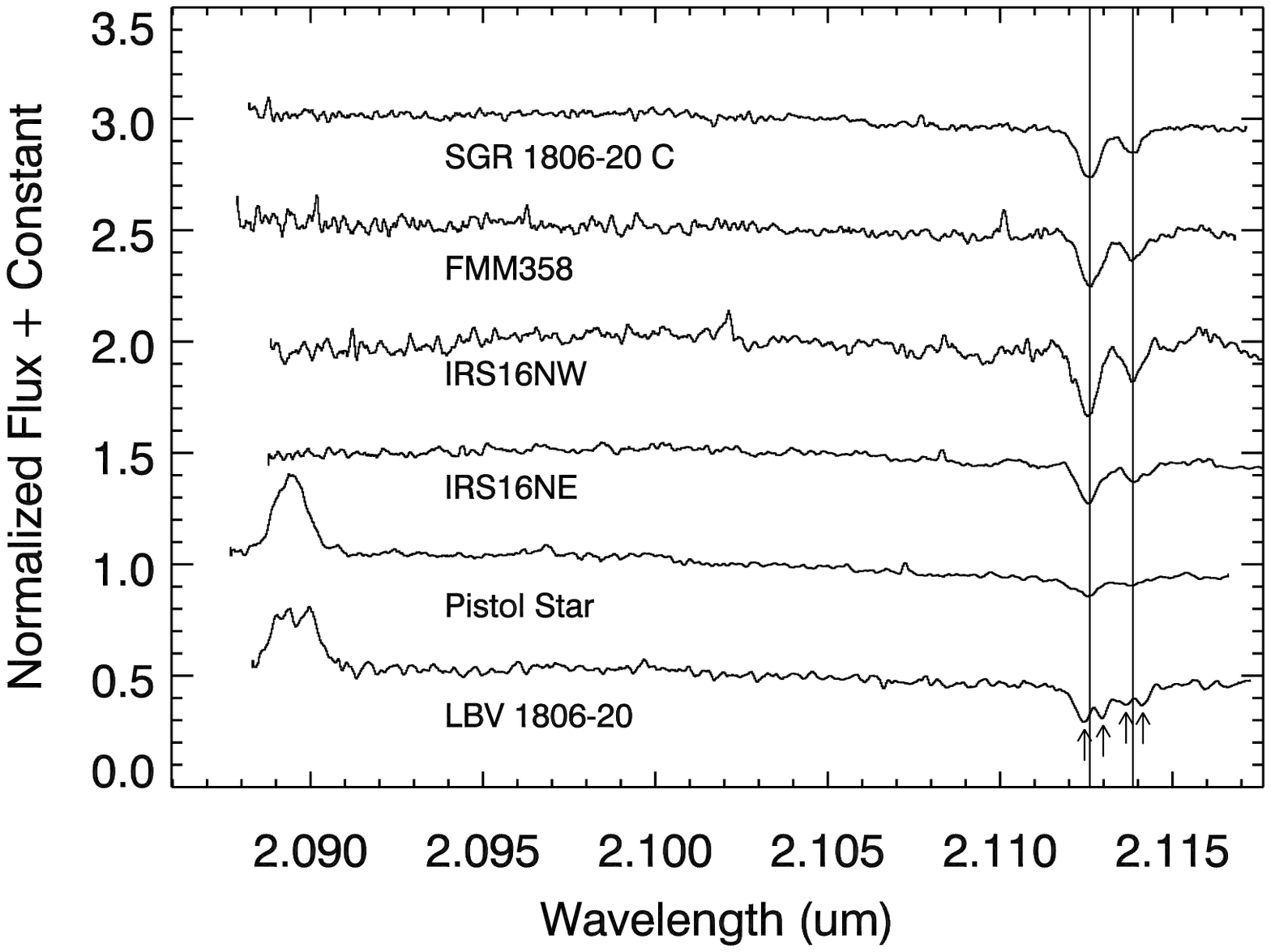}
\caption
{\label{fig:2112}Spectra of massive stars near the 2.112/2.113~\micron\  \ion{He}{1} doublet.
The Pistol Star, IRS16NE, and
\lbv\ are LBVs or LBV-candidates \citep{fig98,tam96,van95}, and the remaining stars are B-type
supergiants \citep{fig99a,gen96}. 
All the spectra show a single
\ion{He}{1} doublet near 2.1125~\micron, except that for \lbv, in which it is obviously double (marked by arrows).
The broad feature near 2.089~\micron\ is common to LBV spectra and corresponds to an \ion{Fe}{2} transition.}
\end{figure}

\begin{figure}
\plotone{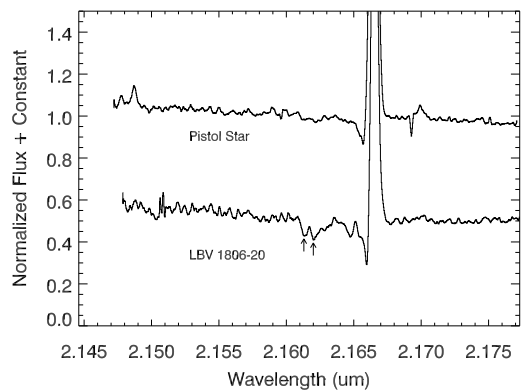}
\caption
{\label{fig:brg}Spectra of the Pistol Star and \lbv\ near the Brackett$-\gamma$ line.
The two absorption lines near 2.160~\micron\ in the spectrum
for \lbv\ (marked by arrows) are due to two separate transitions of \ion{He}{1}.}
\end{figure}

\begin{figure}
\plotone{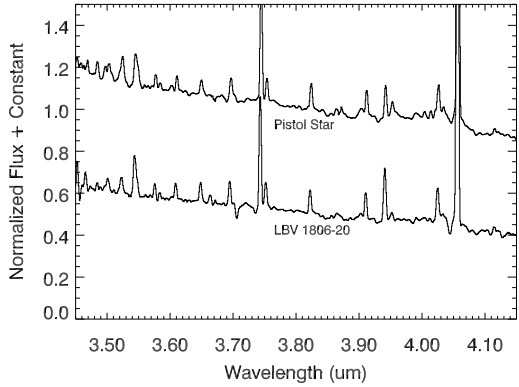}
\caption
{\label{fig:bra}Spectra of the Pistol Star and \lbv\ in the 3-4~\micron\ region. Both spectra
have a strong Brackett$-\alpha$ emission line near 4.05~\micron, a Pfund$-\gamma$ (\ion{H}{1} 6-5) emission line
near 3.74~\micron, and a set of emission lines in the Humphreys series (\ion{H}{1} n-6). 
Hydrogenic transitions of \ion{He}{1} can be seen in absorption just shortward of
the Brackett$-\alpha$ lines. The remaining lines are mostly due to \ion{Fe}{2}.}
\end{figure}

We subtracted background flux, dark current, and residual bias by differencing the nod pair
frames, and we divided the result by the flat field frame in order to compensate
for non-uniform response. We extracted spectra of the objects and applied a
wavelength solution based upon the locations of arc lines. Finally, we divided
the target spectra by the spectrum for Q3 in order to remove telluric effects. Note
that we omitted this last step in the case of the spectra near 1.7~\micron\ because
Q3 is too faint at such short wavelengths. The final extracted spectra are
shown in Figures~\ref{fig:1700}, \ref{fig:2112}, \ref{fig:brg}, and \ref{fig:bra}. They have
been individually shifted along the x-axis so that the spectral features are
aligned with their vacuum wavelengths at rest.

\section{Discussion}

\subsection{Spectral Analysis}

The spectra of \lbv\ and the Pistol Star are virtually indistinguishable, 
except for the morphology of the \ion{Fe}{2} and \ion{He}{1} lines. The latter
are generated in the photospheres and winds of LBVs and are invariably
seen in absorption. In \lbv, the \ion{He}{1} lines are obviously double, with a velocity
separation of 72$\pm3$~\kms. The short-wavelength component of the 1.70~\micron\ line 
has a P Cygni signature, indicating that the star responsible for the bulk of the
emission lines is blue-shifted with respect to the other star. Indeed, none of
the emission lines are double, indicating that the red-shifted component likely
has less luminosity and/or a thinner wind. The velocity separation between the
two minima in the components is 106~\kms, larger than the separation of the components
near 2.112~\micron, as would be expected given the P Cygni nature of the profile in
the blue component of the 1.700~\micron\ line. The \ion{He}{1} absorption component
near 2.160~\micron\ appears to be double (marked by arrows), with a separation of 99~\kms; however, we believe
the lines correspond to two expected \ion{He}{1} lines 
with velocity seperation of 83~\kms\ near that wavelength. 
Figure~\ref{fig:bra} does not appear to show any double absorption lines in the spectrum of \lbv, presumably
due to the relatively low spectral resolution of the data (70~\kms\ corresponds roughly
to 2 pixels in the low resolution mode). The absorption feature near 4.04~\micron\ is
likely due to a blend composed of a \ion{He}{1} doublet (5f$^3$F$_o$-4d$^3$D at 4.0377~\micron\ and 
5f$^1$F$_o$-4d$^1$D at 4.0409~\micron). 
The \ion{He}{1} absorption blend near 3.703~\micron\ (5d$^3$D-4p$^3$P$_o$) has a FWHM of 183~\kms,
compared to a FWHM of 152~\kms\ for the nearby Pfund-$\gamma$ line at 3.74~\micron.

The \ion{Fe}{2} lines are broad in the spectra of both the Pistol star and \lbv, but they have
multiple peaks in the latter. The 1.688~\micron\  and  2.089~\micron\ lines
have nearly identical morphology, each with three peaks in similar intensity ratios and
relative positions in the spectrum of \lbv. 
The Gaussian shape of the \ion{Fe}{2} lines in the spectrum
for the Pistol star is likely caused by recombination closer to photosphere, where the wind velocity
has not reached its constant terminal value. In the spectrum for \lbv, the iron likely recombines further out in the wind so that
emission is dominated by gas at the terminal velocity, thus causing the
flat top shape of the emission line. This explanation for the variation in the \ion{Fe}{2} 
line shapes in the two stars suggests that \lbv\ has slightly higher 
ionizing radiation, consistent with the apparent P-Cygni nature of the blue component
of the \ion{He}{1} lines in the spectrum for that star. 

Spectra of both stars have similar morphology in the H lines. However, the presence of deeper absorption features
in the observed H lines of the \lbv, as well as their narrower nature, are consistent with underlying 
absorption H lines from the putative companion.

We estimate the line-of-sight velocity for \lbv\ by comparing
the vacuum wavelength of the \ion{He}{1} line near 2.1125787~\micron\ with its
measured positions at 2.1125065~\micron\ and 2.1130192~\micron, as determined by
fitting Gaussian profiles to the two absorption components. We adjust the velocity
to the heliocentric frame by adding +0.2~\kms\ to the measured velocity 
to compensate for Earth motion at the time of observations. We add
+$10\pm0.4$~\kms\ to this value in order to adjust the velocity to the local standard of
rest frame \citep{deh98}. Assuming a roughly equal mass binary, we average the velocities of the two absorption lines, 
to find a systemic velocity of V$_{\rm LSR}$=36.4~\kms; note that a change in this
assumption, to a mass ratio of 1:1.5, would only induce a change in our estimate of $\pm$4~\kms. 
Our estimate is similar
to those we obtain for three nearby hot stars, for which V$_{\rm ave, LSR}$=35.0$\pm$5.7~\kms\ (see Figure \ref{fig:2112}).
These velocities contrast with the value in \citet{eik04} of 10$\pm$20~\kms. 

\begin{figure}
\plotone{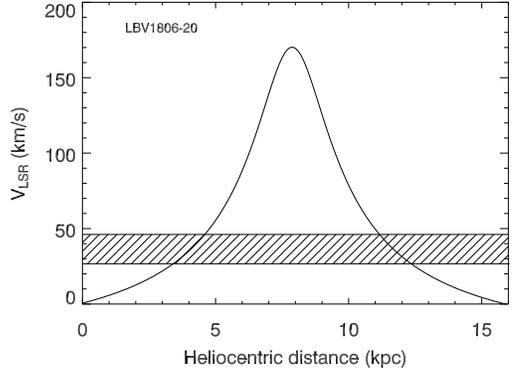}
\caption
{\label{fig:rot}Galactic rotation curve toward the line of sight of \lbv\ from \citet{bra93}. 
The cross-hatched region corresponds to our estimate of the systemic velocity of \lbv\ in the
LSR frame. The error band limits are $\pm$10~\kms, where 6~\kms\ represents the
measurement error, and 4~\kms\ allows for mass ratios between 1:1 and 1:1.5.}
\end{figure}

The bulk velocity of the \lbv\ cluster can be interpreted in the context of the Galactic rotation curve
in \citet{bra93} in order to estimate its distance. We find a distance
of 4.1~kpc or 11.8~kpc (see Figure~\ref{fig:rot}). Only the high solution is consistent 
with a location beyond an NH$_3$ cloud at a distance
of 5.7~kpc \citep{bra93}. This estimate is considerably less than the
estimate in \citet{eik04} of 15.1$^{+1.8}_{-1.3}$~kpc. 

\subsection{Mass of \lbv}
For the far distance estimate (d$\sim$11.8~kpc), and the photometry in \citet{eik04}, we find M$_K$=$-$9.5$\pm$0.1. 
It is difficult to convert this 
value into a luminosity estimate because the range of temperatures, and, therefore
the K-band bolometric correction, varies widely for LBVs \citep{blu95}. However,
if we use the BC$_K$ for the Pistol Star from \citet{fig98}, then we arrive at a luminosity
of 10$^{6.3}$~\Lsun. Using Langer's stellar evolution models in \citet{fig98}, we find an initial mass
$\sim$130~\Msun\ in the case that the object is a single star.

\subsection{\lbv\ as a Massive Binary}
We suggest that the object is a massive binary, based upon the double lines
in its spectrum. For a 1:1 mass system, the components have initial masses of 
65~\Msun, assuming a mass-luminosity relation of M$\propto$L$^{\alpha}$, where alpha$\sim$1.
Our present analysis is fairly insensitive to the mass ratio, a topic that we will
revisit in a future study.
By comparison, the initial masses of the components in $\eta$~Car are 120~\Msun\ and 40~\Msun\
\citep{dam97,pit02}. Another massive binary system, WR20a, contains roughly equal mass
components, each having M$\sim$70-80~\Msun\ \citep{rau04,bon04}.


LBV~1806$-$20 is similar, in appearance, to a growing group of
objects, i.e.\ the Pistol Star, FMM362, $\eta$~Car, and others. Whether
these objects are single or multiple, they all presumably have at least one 
evolved component that is massive ($\simgr$50~\Msun) and some are
surrounded by circumstellar material produced by massive eruptions. 

\subsection{The Formation of the Most Massive Stars}
Theoretical arguments suggest that stars with initial masses $\simgr$150~\Msun\ 
have hydrogen-burning (main sequence) lifetimes of about 2~\Myr\ \citep{bon84}, and
total lifetimes no more than $\sim$3~\Myr\ \citep{mey94}. If \lbv\ is single, and
it had an initial mass in this range, then it would necessarily be much younger than 
nearby Wolf-Rayet stars in the same cluster that are at least 3~\Myr\ old \citep{mey94}. \citet{eik04} suggest that
a supernova from a first generation of stars could have triggered the formation
of \lbv, and perhaps other nearby stars. While this mechanism is often invoked to
explain mutli-generation star formation, it is not the only
explanation in this case. 
Note that the LBVs in the Quintuplet cluster should not live longer than 
$\sim$3~\Myr\ old, whereas the other cluster members are 4~\Myr\ old; 
however, in this case, the original natal cloud must have been strongly
sheared by the tidal field in the Galactic center on time scales much
less than 2~\Myr. A more general explanation for the
age discrepancy in both the Quintuplet and in the \lbv\ cluster
might be that the most massive stars are binary systems of less massive
stars (which can live longer), or that they were binary systems that recently
merged \citep{fig02}. 

This proposition also resolves the problem of explaining why 200~\Msun\ stars
are observed in the Quintuplet cluster, whereas the upper mass cutoff to
the initial mass function in the nearby Arches cluster is 
$\sim$150~\Msun\ \citep{fig03}. In fact, if single stars
with masses of $\sim$100~\Msun\ exist, then where are the binary systems
with such stars? Certainly, some of the most massive stars exist in
binary systems, as evidenced by the substantial binary fraction (~25\%) in the 
WR catalogue \citep{van01}. Perhaps we are witnessing the ancestors of such systems 
in these LBVs. 



The evidence in this paper suggests that \lbv\ may be binary, but we will
continue to explore the possibility that the double-lined spectra could have
an alternate explanation. For instance, perhaps the profile is a composite of
an emission feature, generated by a disk or wind, at the systemic
velocity and superposed on a photospheric absorption feature. In addition, we
will refine the temperature, luminosity, and mass ratio estimates through a more detailed 
analysis, coupled with wind/atmosphere
models \citet{naj04}. In addition, we have begun a spectroscopic
monitoring program to detect changes that might
indicate a binary nature for \lbv.

\acknowledgements
F. N. acknowledges grants AYA2003-02785-E and ESP2002-01627.
We thank Keck staff members Randy Campbell and Grant Hill. 
We acknowledge useful discussions with Richard Larson, Stephen Eikenberry, and
Augusto Damineli.

\end{document}